\documentclass{llncs}

\usepackage{natbib}
\usepackage{graphicx}

\usepackage{fancyhdr}
\usepackage{lastpage}
 
\begin{document}

\title{A short review and primer on respiration in human computer interaction applications}
\author{Ilkka Kosunen\inst{1} \and Benjamin Cowley\inst{2,3}}
\institute{Helsinki Institute for Information Technology HIIT, Department of Computer Science, University of Helsinki, Helsinki, Finland
\and
Finnish Institute of Occupational Health,\\
\email{benjamin.cowley@ttl.fi},\\
POBox 40, Helsinki, 00250, Finland
\and
Cognitive Brain Research Unit, Institute of Behavioural Sciences, University of Helsinki, Helsinki, Finland}

\maketitle              

\begin{abstract}
The application of psychophysiology in human-computer interaction is a growing field with significant potential for future smart personalised systems. Working in this emerging field requires comprehension of an array of physiological signals and analysis techniques. 

Respiration is unique among physiological signals in that it can be consciously controlled which has to be taken into account when designing applications. Respiration is tightly connected to other physiological signals, especially cardiovascular activity, and often analyzed in conjunction with other signals. When analyzed separately, an increase in the rate of respiration can be seen as an increase in metabolic demand which indicates activate states such as engagement and attention. We present a short review on the application of respiration in human-computer interaction. 

This paper aims to serve as a primer for the novice, enabling rapid familiarisation with the latest core concepts. We put special emphasis on everyday human-computer interface applications to distinguish from the more common clinical or sports uses of psychophysiology.

This paper is an extract from a comprehensive review of the entire field of ambulatory psychophysiology, including 12 similar chapters, plus application guidelines and systematic review. Thus any citation should be made using the following reference:

{\parshape 1 2cm \dimexpr\linewidth-1cm\relax
B. Cowley, M. Filetti, K. Lukander, J. Torniainen, A. Henelius, L. Ahonen, O. Barral, I. Kosunen, T. Valtonen, M. Huotilainen, N. Ravaja, G. Jacucci. \textit{The Psychophysiology Primer: a guide to methods and a broad review with a focus on human-computer interaction}. Foundations and Trends in Human-Computer Interaction, vol. 9, no. 3-4, pp. 150--307, 2016.
\par}


\end{abstract}

\section{Introduction}


The respiration signal is of scientific relevance because it is controlled by the autonomic nervous system and the central nervous system. This means that respiration can be consciously controlled by the subject (unlike, for instance, heart rate). Different parts of the brain are responsible for autonomic and behavioural breathing: voluntary breathing is controlled by the cerebral cortex, while autonomous breathing is controlled by the brain stem. There are, however, projections from the cortex to the brain stem that allow the higher centres to influence metabolic breathing patterns. The autonomic respiratory pattern is therefore a complex interaction among the brain stem, the limbic system, and the cortex \citep{Homma2008}.

While respiration is an interesting signal from a theoretical perspective, there are some practical issues that limit its use in an HCI setting. Traditional respiratory measurement techniques, such as spirometers, which force the user to breathe through a tube, are often overly intrusive and cumbersome, while non-invasive methods can be too imprecise. The fact that respiration can be consciously controlled can act as a confounding factor. Also, respiration is highly susceptible to artefacts produced by speech and movement \citep{wientjes1992}.

\section{Background}
Although respiration (respiration rate) is an important topic in health science, in physiology studies in a games or learning context it is less well researched than are many other psychophysiological signals, suffering from neglect even in studies of the cardiovascular system, with which it interacts intimately \citep{Nesterov2005}. This is partly an interpretative issue, as respiratory control is both voluntary and involuntary \citep{Harver2000}, but equally it is a function of the uses served by respiration -- the heart and brain together account for less than 3\% of body weight but more than 30\% of oxygen usage. Accordingly, states of low mental workload and high metabolic rate may be reflected in the respiration rate similarly to states of high mental workload and low metabolic rate. However, it has been shown that increased respiration rate results when subjects are sitting and attentively wakeful (e.g., listening to a story), as opposed to sitting without paying attention or with closed eyes. Thus, in similar metabolic situations, increased respiration indicates an increase in attention \citep{Harver2000}.
In consequence, given an attentionally demanding situation with static metabolic demand, respiration should increase in line with the subject's engagement. In other words, the more attention paid (barring metabolic variation), the higher the respiration rate.



In its effects, respiration is closely interwoven with most other physiological signals of interest, because it provides the oxygen that underlies localised mental activity, the energising of the musculature, and heart rate response \citep{Harver2000}. Respiration affects heart rate variability through a process called `respiratory sinus arrhythmia' (RSA) whereby the inter-beat interval of the heart is shortened during inspiration and prolonged during expiration \citep{Yasuma2004}. It may therefore be necessary to control for breathing in experiments that include HRV.

In addition to the respiratory rate, there are several (more specific) indices that can be derived from the respiratory pattern, such as inspiratory time, inspiratory volume, and inspiratory pause. A study by \cite{Boiten1998} explored the effect of film scenes, reaction time, and cold pressor tasks on affect-related respiratory patterns. The study showed that there were clear effects on respiratory patterns whenever the films elicited amusement or disgust, in that amusement induced a decrease in inspiratory time and tidal volume, while disgust would elicit prolonged inspiratory pauses (breath-holding). The reaction time task induced a relatively fast, shallow, and regular breathing pattern, while the cold pressor brought on a substantial increase in respiratory volume, an increase in post-inspiratory pause duration, and a large amount of breath-to-breath variability in the pattern of breathing. Accordingly, in the design of HCI applications it can be useful to pay attention to more specific details of respiration than respiration rate alone (see \citet{Kreibig2010} for a comprehensive list of possible respiratory pattern indices).

Frequent sighing has been shown to be associated with anxiety. \cite{blechert2006identifying}, for instance, studied the physiological correlates of anxiety and found that sighing frequency increased 150$\%$ when the participants thought that they might receive an electric shock.

\section{Methods}
\begin{figure}[!ht]
	\centering
	\includegraphics[scale=0.5]{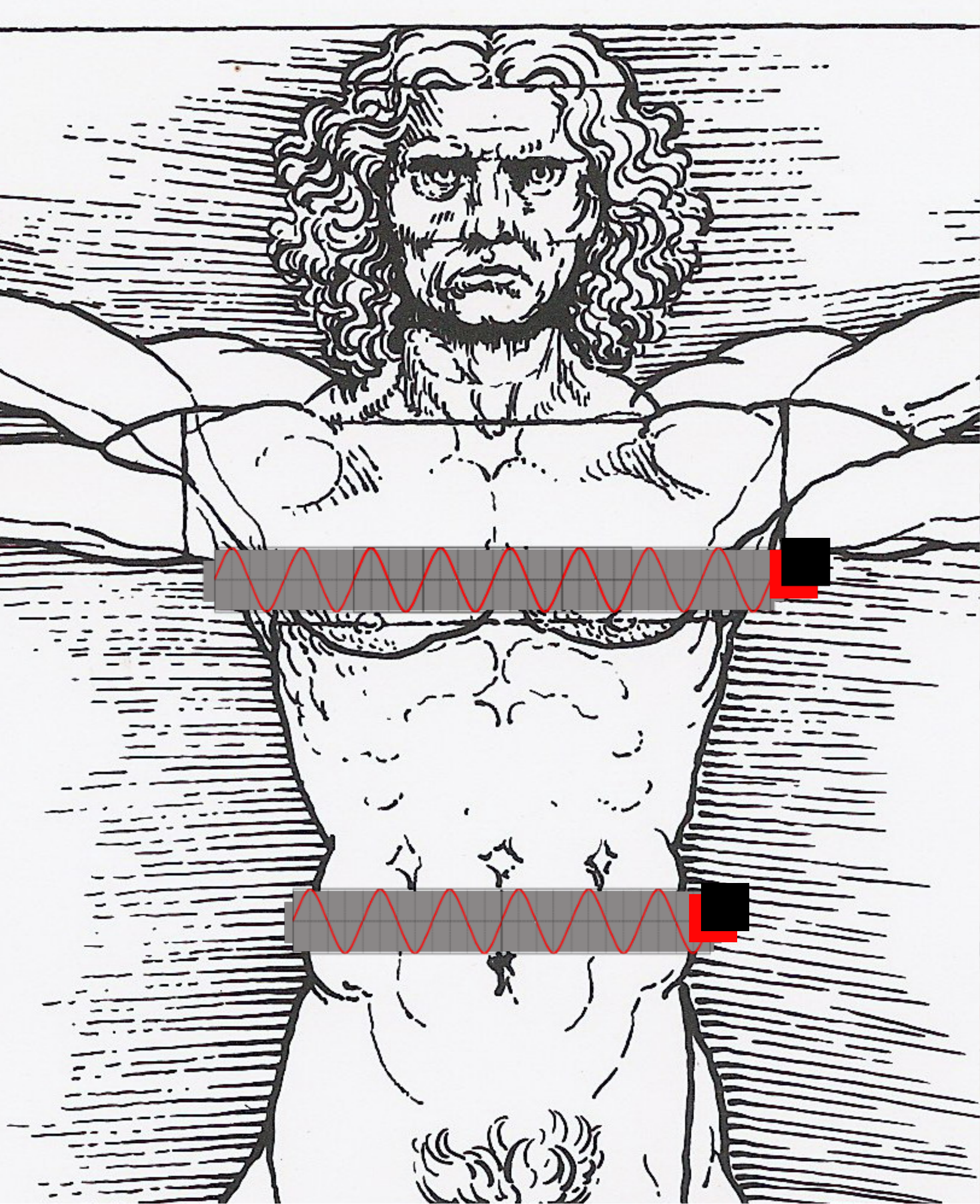}
	\caption{Schematic diagram of the canonical locations for respiration belt sensors, on the chest and at the abdomen. Respiration belts are typically a flexible cable transducer (pictured here as a red sinusoid) attached to an elasticated cloth belt (here, a grey rectangle), terminating in connectors to the amplifier (represented by red and black squares).}
	\label{fig.resp_belts}
\end{figure}

Several methods exist for measuring respiration. The flow of air in and out of the lungs can be measured directly by means of a spirometer. However, while very accurate, this apparatus forces the user to wear a nose clip and to breathe through a tube, which makes it unsuitable for most HCI purposes. Therefore, instead of measuring respiration directly, it is often more convenient to measure the movement of the chest and the abdomen, by using respiratory inductance plethysmography (RIP). In this method, an elastic band is wrapped around the abdomen and around the chest. It is also possible to use only a single belt, around the abdomen. However, to estimate tidal volume (i.e., the total volume of air inspired and expired with each breath), it is necessary to use two belts, enabling both the abdomen and the chest to be measured (see Figure~\ref{fig.resp_belts}). Nonetheless, because of the physiological uniqueness of every individual, the measurement of tidal volume with two belts requires calibration of the system for each user, through use of a spirometer \citep{wientjes1992}.

\section{Applications}
In an HCI context, respiration has been used mostly as an explicit control signal. ~\citet{Moraveji2011} studied whether they could influence the respiration of users during an information-processing task. 
Reducing breathing rate has been shown to reduce stress and anxiety, so the aim was to build a system that helps users to regulate their breathing to this end. It displayed a semi-translucent bar in the bottom half of the screen that served as guidance for the user. Users were instructed to inhale when the bar goes up and exhale as it moves down. The system was designed to illustrate a breath rate 20$\%$ below the user's baseline. It indeed reduced users' breathing rate -- the authors found that users were able to decrease their breathing rate significantly while still able to continue performing information work such as research, writing, and programming.

The use of respiration in a more entertainment-oriented context was studied by ~\citet{Kuikkaniemi2010}, who explored the use of biofeedback in computer gaming. They developed a first-person shooter game wherein the fire rate and recoil were linked to the player's respiration. In their experiment, they studied how biofeedback was perceived when the players were aware of the adaptation as compared to when they were unaware of the biofeedback. The researchers established that conscious control through respiration was more immersive and rewarding, thereby highlighting the nature of respiration as a signal that can be dealt with as either an implicit or an explicit form of interaction.

\section{Conclusions}
Respiration is a signal that is rarely used on its own. More often, is serves as an auxiliary signal either to complement information from other signals, such as heart rate in the case of respiratory sinus arrhythmia, or to serve as part of a machine learning approach that draws together information from several physiological signals. When used alone, respiration is usually employed as an explicit control mechanism, offering an additional input channel that can complement manual input.

\bibliographystyle{plainnat}
\bibliography{ch3_respiration_bib}

\end{document}